\documentclass[aps,prx,twocolumn,superscriptaddress,showpacs,longbibliography]{revtex4-1}

\usepackage{graphicx}
\usepackage{dcolumn}
\usepackage{bm}
\usepackage{amssymb}
\usepackage{amsmath}
\usepackage{color}
\usepackage{natbib}

\newcommand{\DSO}{DyScO$_3$}
\newcommand{\DFO}{DyFeO$_3$}
\newcommand{\mB}{$\mu_{\rm{B}}$}

\begin{document}

\title{Magnetic ground state of the Ising-like antiferromagnet \DSO}

\author{L.~S.~Wu}
\affiliation{Quantum Condensed Matter Division, Oak Ridge National Laboratory, Oak Ridge, Tennessee 37831, USA}
\author{S.~E.~Nikitin}
\affiliation{Max Planck Institute for Chemical Physics of Solids, N\"{o}thnitzer Str. 40, 01187 Dresden, Germany}
\affiliation{Institut f\"ur Festk\"orper- und Materialphysik, Technische Universit\"at Dresden, D-01069 Dresden, Germany}
\author{M.~Frontzek}
\affiliation{Quantum Condensed Matter Division, Oak Ridge National Laboratory, Oak Ridge, Tennessee 37831, USA}
\author{A.~I.~Kolesnikov}
\affiliation{Chemical and Engineering Materials Division, Oak Ridge National Laboratory, Oak Ridge, Tennessee 37831, USA}
\author{G.~Ehlers}
\affiliation{Quantum Condensed Matter Division, Oak Ridge National Laboratory, Oak Ridge, Tennessee 37831, USA}
\author{M.~D.~Lumsden}
\affiliation{Quantum Condensed Matter Division, Oak Ridge National Laboratory, Oak Ridge, Tennessee 37831, USA}
\author{K.~A.~Shaykhutdinov}
\affiliation{Kirensky Institute of Physics, Federal Research Center SB RAS, Krasnoyarsk, 660036 Russia}
\author{E.-J.~Guo}
\affiliation{Quantum Condensed Matter Division, Oak Ridge National Laboratory, Oak Ridge, Tennessee 37831, USA}
\author{ A.~T.~Savici}
\affiliation{Neutron Data Analysis and Visualization Division, Oak Ridge National Laboratory, Oak Ridge, Tennessee 37831, USA}
\author{ Z.~Gai}
\affiliation{Center for Nanophase Materials Sciences and Chemical Science Division, Oak Ridge National Laboratory, Oak Ridge, Tennessee 37831, USA}
\author{ A.~S.~Sefat}
\affiliation{Materials Science and Technology Division, Oak Ridge National Laboratory, Oak Ridge, Tennessee 37831, USA}
\author{A.~Podlesnyak}
\thanks{Corresponding author. Electronic address: podlesnyakaa@ornl.gov}
\affiliation{Quantum Condensed Matter Division, Oak Ridge National Laboratory, Oak Ridge, Tennessee 37831, USA}
\date{\today}

\begin{abstract}
We report the low temperature magnetic properties of the \DSO\ perovskite, which were characterized by means of single crystal and powder neutron scattering, and by  magnetization measurements.
Below $T_{\mathrm{N}}=3.15$~K, Dy$^{3+}$ moments form an antiferromagnetic structure with an easy axis of magnetization lying in the $ab$-plane.
The magnetic moments are inclined at an angle of $\sim\pm{28}^{\circ}$ to the $b$-axis.
We show that the ground state Kramers doublet of Dy$^{3+}$ is made up of primarily $|\pm 15/2\rangle$ eigenvectors and well separated by crystal field from the first excited state at $E_1=24.9$~meV.
This leads to an extreme Ising single-ion anisotropy, $M_{\perp}/M_{\|}\sim{0.05}$.
The transverse magnetic fluctuations, which are proportional to $M^{2}_{\perp}/M^{2}_{\|}$, are suppressed and only moment fluctuations along the local Ising direction are allowed.
We also found that the Dy-Dy dipolar interactions along the crystallographic $c$-axis are 2-4 times larger than in-plane interactions.
\end{abstract}


\maketitle

\section{Introduction}
\label{Intr}
Orthorhombic \DSO\ is a member of the rare-earth perovskite family $RM$O$_3$ where $R$ is a rare-earth ion and $M$ is a transition metal ion.
Magnetic properties of these compounds have attracted continued attention due to a number of intriguing physical phenomena, like ferroelectric \cite{Cohen,Lee} and multiferroic \cite{Cheong,Khomskii} properties, temperature and field induced spin reorientation transitions \cite{Belov}, magneto-optical effects \cite{Kimel}, or exotic quantum states at low temperatures \cite{Mourigal}.
In orthoperovskites with $M=$~Fe, Mn, the sublattice of $3d$ moments typically undergoes an ordering transition at several hundreds of Kelvin, whereas the $4f$ sublattice only orders at a few Kelvin, indicating much stronger exchange coupling between the $3d$ moments~\cite{White}.
The interaction between the two spin subsystems, however, also plays an important role and often determines the magnetic ground state.
For instance, the interplay between the Fe and Dy sublattices gives rise to gigantic magnetoelectric phenomena in \DFO~\cite{Tokunaga}.
In the case of non-magnetic $M=$~Al, Sc, or Co (in its low-spin state), the magnetic properties of $RM$O$_3$ are primarily controlled by an electronic structure of $R^{3+}$ ion and rare-earth inter-site interactions.
In turn, the crystalline electrical field (CEF) splitting of the lowest lying $4f$ free-ion state determines the single ion anisotropy as well as the magnitude of the magnetic moment.

In spite of its three-dimensional perovskite structure, \DSO\ has been reported as a highly anisotropic magnetic system with an antiferromagnetic (AF) transition, $T_{\mathbf{N}} \simeq{3.1}$~K~\cite{Ke,Raekers,Bluschke}.
The details of the \DSO\ magnetic state and the Dy-Dy interactions at low temperatures remained poorly understood.
Recent studies of \DSO\ lead to conflicting conclusions on the magnetic anisotropy and ground state of rare-earth subsystem.
The compound exhibits strong magnetic anisotropy with moments confined in the $ab$-plane.
On the one hand, it was suggested that an easy axis is along the $a$-axis~\cite{Ke}.
Indeed, such spin configurations were found in some related isostructural compounds, such as YbAlO$_3$~\cite{Radha81},  TbCoO$_3$~\cite{Knizek}, and SmCoO$_3$~\cite{Jirak}.
On the other hand, recent magnetization measurements of \DSO\ suggest that an easy axis is along the crystallographic $b$ direction~\cite{Bluschke}.
Besides, in other perovskites with $R=$~Dy, the easy axis of magnetization was reported to be along the $b$-axis, for example DyCoO$_3$~\cite{Knizek} and DyAlO$_3$~\cite{Schuchert}.
Dy$^{3+}$ in electronic configuration $4f^9$ is a Kramers ion split by the crystal electric field (CEF), resulting in eight doublets.
Since the CEF is controlled by the near neighbor coordination which is little affected by an isostructural substitution of ligands, the different ground state of \DSO\ compared to other Dy$M$O$_3$ compounds looks puzzling.

In this article we use the neutron scattering and magnetization measurements to study the magnetic ground state of \DSO\ in more detail.
We show that the magnetic properties at low temperatures are dominated by the ground state Kramers doublet $\lvert\pm{15}/2\rangle$, which is characterized by an Ising single-ion anisotropy and moments fluctuating \emph{along} the local Ising direction.
We find that the Dy$^{3+}$ ordered moments are canted $28^{\circ}$ away from the $b$-axis, which is in agreement with other Dy$M$O$_3$ isostructural perovskites.

\section{Experimental Details}

In this work we used high quality single crystals of \DSO, which are commercially available because they are commonly used as substrates for epitaxial ferroelectric and multiferroic perovskite thin film growth~\cite{Choi,Schlom}.
For magnetic measurements \DSO\ crystals were oriented using an x-ray Laue machine and then cut with a wire saw to get planes perpendicular to the $a$, $b$ and $c$-axes.
From the Laue patterns we estimate the orientation to be within $\sim{1}^{\circ}$.
Magnetization was measured using a vibrating sample SQUID magnetometer (MPMS SQUID VSM, Quantum design) in the temperature range $2-300$~K.

Neutron powder diffraction (NPD) was measured with a crushed single crystal of a total mass $\sim{0.5}$~g, at the wide-angle neutron diffractometer WAND (HB-2C) at the HFIR reactor at Oak Ridge National Laboratory (ORNL).
The sample was enclosed in a hollow Al cylindrical sample holder, in order to diminish the strong neutron absorption by Dy, and placed into a He flow cryostat to achieve a minimum temperature of $T={1.5}$~K.
An incident neutron beam with a wavelength of 1.4827~{\AA}~ was selected with a Ge (113) monochromator.

High energy transfer inelastic neutron scattering (INS) experiments were performed at the Fine Resolution Fermi Chopper Spectrometer (SEQUOIA) at the Spallation Neutron Source (SNS) at ORNL~\cite{SEQ1,SEQ2}, using the same powder sample.
The data were collected at $T={6}$~K with an incident neutron energy of $E_{\rm{i}}=100$~meV, resulting in an energy resolution of Gaussian shape with full width at half maximum (FWHM) $\sim{3}$~meV at the elastic line.

The single crystal quasielastic mapping and low energy transfer INS measurements were done at the Cold Neutron Chopper Spectrometer (CNCS) \cite{CNCS1,CNCS2}.
A bar shaped ($0.4\times{4}\times{20}$~mm$^3$) single crystal of mass $\sim{0.2}$~g was oriented in the $(H0L)$ scattering plane.
The detector coverage out-of-plane was about $\pm{15}^{\circ}$, so that a limited $Q$ range along the $K$-direction could also be accessed.
The data were collected using a fixed incident neutron energy of 3.2~meV.
In this configuration, the energy resolution was $\sim{0.07}$~meV (FWHM) at the elastic line.

Data reduction and analysis was done with the \textsc{Fullprof}~\cite{FP}, \textsc{SARA}{\it{h}}~\cite{Wills}, \textsc{MantidPlot}~\cite{Mantid} and \textsc{Dave}~\cite{Dave} software packages.

\section{Results and Analysis}
\subsection{Crystal structure and crystal electric field}
\label{sect_CEF}

\begin{figure}[tb]
\includegraphics[width=1\linewidth]{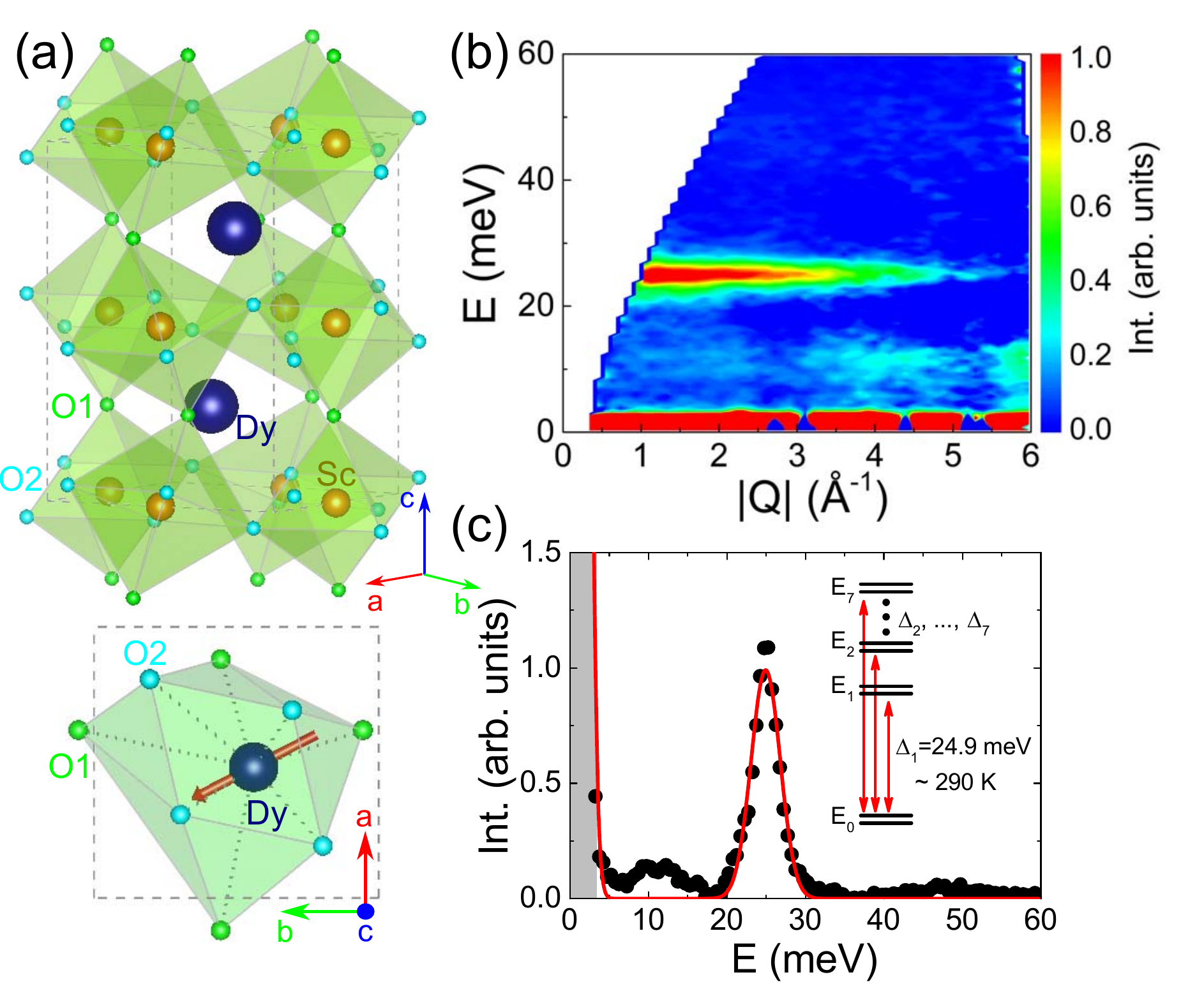}
\caption{(a) Top: Crystal structure of \DSO, where a Dysprosium (Dy) atom is surrounded by eight distorted Scandium-Oxygen (Sc-O) octahedra. Bottom: Local environment of Dy$^{3+}$ in the $z=1/4$ plane, considering twelve nearest Oxygen (O) neighbors (four O1 sites in the same $z=1/4$ plane, four O2 sites above and four O2 sites below the $z=1/4$ plane). The red arrows indicate the Dy$^{3+}$ Ising moment direction. (b) Contour plot of the INS spectrum of \DSO\ taken at the SEQUOIA spectrometer at temperature $T=6$~K. The excitation to the first CEF level was observed centering around 25~meV. The scattering intensity decreases with increasing wave vector $|Q|$, indicating the magnetic nature of the transition due to the form factor. (c) Plot of the integrated intensity over wave vector $|Q|=[1,3]$~\AA$^{-1}$~ as a function of the energy transfer $E$. The first excited CEF level was fitted with a Gaussian function, which peaks at $\Delta_{1}=24.93\pm{0.02}$~meV, as indicated by the red solid line. The gray bar area indicates the instrumental resolution. Inset: Sketch of the eight isolated CEF doublet states ($E_1, E_2,...,E_7$) of Dy$^{3+}$, due to the low local site symmetry, with the ground doublet well separated from all other levels.}
\label{CEF}
\end{figure}

\DSO\ crystallizes in a distorted orthorhombic perovskite structure ~\cite{Liferovich,UECKER}.
Lattice parameters refined from our powder neutron diffraction data at 10~K are $a={5.4136}(3)$~{\AA}, $b={5.6690}(1)$~{\AA}, and $c={7.8515}(1)$~\AA,  using the conventional $Pbnm$ space group (lattice parameters $a<b<c$).
The crystal structure of \DSO\ is illustrated in Fig.~\ref{CEF}(a), where Dysprosium (Dy) atoms are located between eight distorted Scandium-Oxygen (Sc-O) octahedra.
Due to the distortion, the point symmetry of the Dy site is lowered to $C_{\rm{s}}$, with only one mirror plane normal to the $c$-axis.
Therefore, Dy$^{3+}$ moments as constrained by the point symmetry would either point along the $c$-axis, or lie in the $ab$-plane.
To have a quantitative description of the CEF effect, calculations based on the point charge model~\cite{Stevens,Hutchings} were performed using the software package McPhase~\cite{McPhase}.
The first twelve nearest Oxygen (O) neighbors around the Dy ion were considered (bottom of Fig.~\ref{CEF}(a)), which keep the correct local point symmetry of the Dy site and thus constrain the CEF wave functions.
In this local environment, the sixteen fold degenerate $J=15/2$ ($L=5$, $S=5/2$) multiplet ($2J+1=15$) of Dy$^{3+}$ is split into eight doublet states.
By local point symmetry, no high-symmetry directions in the $ab$ mirror plane are present. Therefore, the resulting Ising axis is tilted away from the crystal $a$ and $b$ axes.
The tilting angle $\varphi$ is determined by the relative distortion of the nearest oxygen octahedra.
We transform the old basis ($x, y, z$) along the principle crystal ($a, b, c$) to the new basis ($x', y', z'$), with $\varphi$ defined as the titling angle from the crystal $b$ axis:
\begin{align*}
       &\mbox{old\ basis}  & &\mbox{new\ basis}\\
       x&=(1, 0, 0),  & x'&=(-\rm cos\varphi, sin\varphi, 0)\\
       y&=(0, 1, 0), & y'&=(0, 0, 1)\\
       z&=(0, 0, 1), &z'&=(\rm sin\varphi, cos\varphi, 0)
\end{align*}
The calculated ground doublet states are best diagonalized when the local $z'$ axis is chosen along the Ising moment easy axis, in which case no imaginary coefficients are left in the ground state wave functions.
The tilting angle can be determined with the CEF calculation, by checking the ground state wave function with different values of angle $\varphi$, which results in $\varphi\sim{25}^{\circ}$. The first two Kramers ground state wave functions are given by
\begin{multline*}
E_{0\pm} = 0.991\lvert\pm{15}/2\rangle
\mp{0.107}\lvert\pm{13}/2\rangle\\
{-0.081}\lvert\pm{11}/2\rangle
\pm{0.014}\lvert\pm{9}/2\rangle
{-0.004}\lvert\pm{7}/2\rangle\\
\mp{0.007}\lvert\pm{5}/2\rangle
{-0.002}\lvert\pm{3}/2\rangle.
\label{E0}
\end{multline*}
The calculated excited levels are located at energies 22.9~meV, 37.6~meV, 43.1~meV, 52.2~meV, 63.3~meV, 79.7~meV, and 98.1~meV.
This calculated CEF scheme indicates a well separated ground state doublet, which is almost entirely made up by the wave function $\lvert\pm{15}/2\rangle$.
The calculated CEF parameters and matrix elements for transitions between these CEF levels are shown in detail in Appendix~\ref{append}.
The calculated wave functions for the first excited doublet are mostly from
$\lvert\pm{13}/2\rangle$, and thus the matrix element for excitations between the ground state and the first excited state is large.
However, for higher excited levels, the wave functions are mostly contributed from
$\lvert\pm{11}/2\rangle$, $\lvert\pm{9}/2\rangle$, ..., $\lvert\pm{3}/2\rangle$,
and the matrix elements between the ground state to these higher excited levels are very weak because of the $\Delta{S}=1$ selection rule.
Thus, the CEF calculations indicate that only the one excitation to the first excited doublet can be observed at low temperatures by INS, when only the ground state is populated.

The CEF calculations based on the point charge model are well consistent with our INS data.
The INS spectrum taken with neutron incident energy $E_{\rm{i}}=100$~meV at $T=6$~K exhibits only one dispersionless excitation as expected, see Fig.~\ref{CEF}(b).
The scattering intensity of this excitation decreases with increasing wave vector $|Q|$,  confirming its magnetic origin.
The intensity integrated over a wave vector range $|Q|=[1,3]$~{\AA}$^{-1}$ is shown in Fig.~\ref{CEF}(b) as function of the energy transfer $E$. By fitting the peak with a Gaussian function (red solid line), the first excited level was determined to be $\Delta_{1}=24.93\pm{0.02}$~meV, which is very close to the calculated value.

Since the ground state is well separated from the other excited CEF levels, the low temperature ($T<\Delta_{1}\simeq{290}$ K) magnetic properties are dominated by the ground state doublet symmetry.
Thus, we could use the low temperature magnetization results to examine the calculated ground state wave function, as will be discussed below in the next section.

\subsection{Magnetization}

\begin{figure}[tb]
  \includegraphics[width=0.7\linewidth]{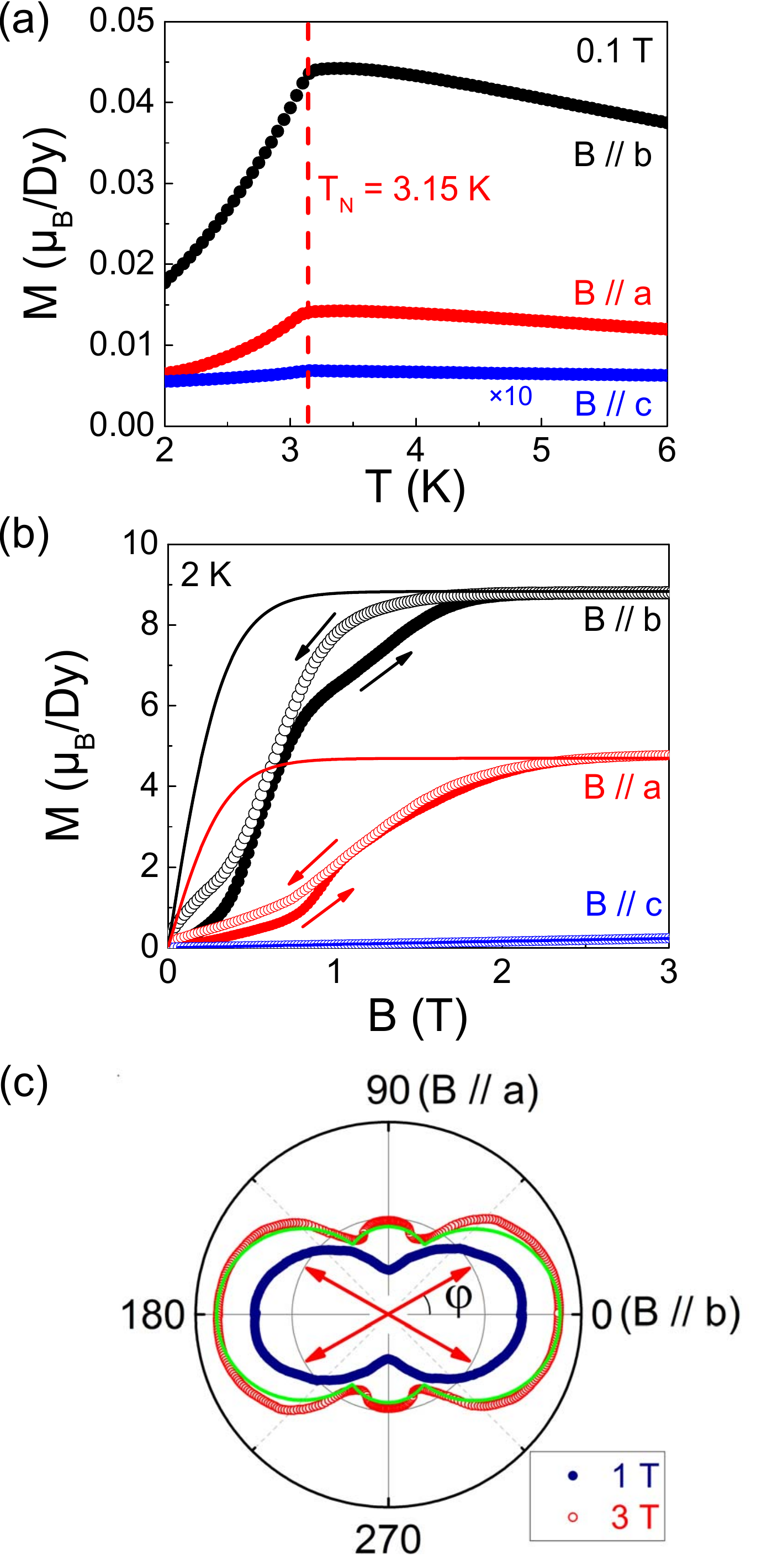}
  \caption{(a) Temperature dependent magnetization of a \DSO\ single crystal with applied field along the $a$, $b$ and $c$-axes. The red dashed line indicates the AF transition at $T_{\rm{N}}=3.15(5)$ K. The magnetization along $c$ was multiplied by a factor 10 for clarity. (b) The field dependent magnetization measured at $T=2$~K. The solid lines are the calculated Brillouin function at 2~K, as explained in the text. (c) Angle dependent magnetization measured at $T=2$~K and field $B=1$~T (blue dots) and $B=3$~T (red dots). The magnetic field was applied in the $ab$-plane. Angle $0^{\circ}$ and $90^{\circ}$ correspond to $B\parallel{b}$ and $B\parallel{a}$, respectively. The green line represents a fit of the experimental data (see main text). Red arrows schematically show a moment configuration of Dy$^{3+}$ at zero field with angle $\varphi=28^{\circ}$ between the $b$-axis and the Dy magnetic moment.}
  \label{MBTA}
\end{figure}

The temperature dependent magnetization of \DSO\ is shown in Fig.~\ref{MBTA}(a),
with the magnetic field applied along the three principle crystallographic axes.
A cusp-like anomaly is observed at $T_{\rm{N}}=3.15(5)$~K for all three directions,
indicating the ordering to an antiferromagnetic phase, that is consistent with previous
reports~\cite{Ke,Raekers,Bluschke}.
The field dependent magnetization measured in the ordered state exhibits a step-like anomaly when the field is applied in the $ab$-plane, see Fig.~\ref{MBTA}(b).
The large observed hysteresis indicates subsequent field-induced first order transitions.
Significant anisotropy was observed between the $ab$-plane and the $c$-axis.
The high field saturation moments along $a$ and $b$ are more than 20 times larger than the moment along the $c$-axis, confirming that the Dy$^{3+}$ magnetic moments are in the $ab$-plane.

To study further the details of the anisotropy, the magnetization was measured in a magnetic field of $B=1$~T and $B=3$~T, with varying direction of the field in the plane.
These measurements are shown in Fig.~\ref{MBTA}(c).
A small hump-like feature appears in the magnetization when the applied magnetic field is large enough to polarize the magnetic moments of both Dy sublattices.
This is an indication that the moment is tilted relative to the $a$ and $b$-axes, as suggested already by the CEF calculations.
The angle dependence of the magnetization in a field $B=3$~T can be well described as:
\begin{align}
M&=\dfrac{M_{\|}}{2}\left(\lvert\cos\left(\theta-\varphi\right)\rvert
+\lvert\cos\left(\theta+\varphi\right)\rvert\right) ,
\end{align}
where $M_{\|}$ is the saturation moment, $\theta$ is the angle between the applied field and the $b$-axis, and $\varphi$ is the angle between the $b$-axis and the Dy$^{3+}$ moment direction.
The result of this analysis is shown as the green line in Fig.~\ref{MBTA}c, with $M_{\|}={10}$~\mB/Dy, and $\varphi={28}^{\circ}$, in good agreement with
ref.~\onlinecite{Bluschke}.
Since the experimental temperature ($T=2$~K) and magnetic field ($B=3$ T) are much smaller than the energy scale of the first excited CEF level $\Delta_{1}$, it is safe to calculate the field dependent magnetization with the Brillouin function of the ground state doublets, Fig.~\ref{MBTA}(b).
While the saturation at high fields can be well described with a saturation moment $M_{\|}={10}$~\mB/Dy, and an angle $\varphi=28^{\circ}$, a large mismatch between the measurements and calculations remains at low magnetic field.
This suggests additional AF correlations in the system, which are missing in the calculation of the Brillouin function.
The magnetization along the $c$-axis can be described with an effective moment $M_{\perp}\simeq{0.5}$~\mB/Dy which is consistent with our CEF calculations.

The agreement between the CEF calculations and the magnetization measurements confirms the Ising character of the ground state wave function $E_{\rm{0}\pm}$.
The very small magnetization along the $c$-axis also shows that the moments directions are strictly constrained.
A direct consequence is that any transverse excitation (spin wave) should be strongly suppressed in \DSO.
The Ising moments are only allowed to fluctuate along their local easy axis, which is reflected in the neutron scattering polarization factor (details to follow below in Section \ref{sec_diffuse}).

\subsection{Neutron powder diffraction and magnetic structure}
\begin{figure}[tb]
  \includegraphics[width=0.9\linewidth]{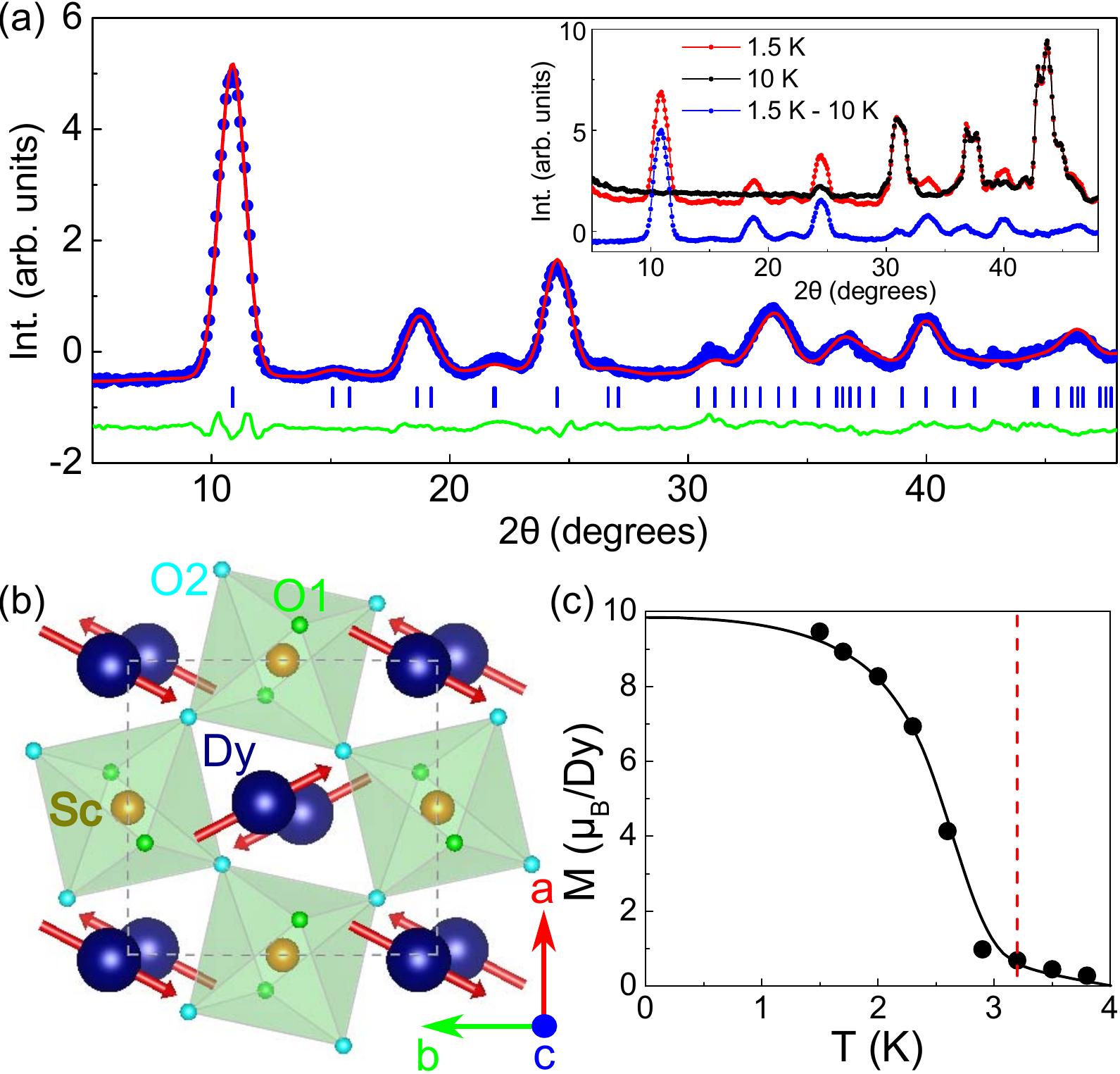}
  \caption{(a) Observed (circles) and calculated (solid line) magnetic NPD patterns for \DSO.
  Bars (blue) show the positions of allowed magnetic reflections. The difference curve (green) is plotted at the bottom. Inset: Raw data taken in the AF ordered state at 1.5~K (red), the paramagnetic state at 10~K (black), and the difference (blue) that shows the magnetic contribution to the scattering. (b) Schematic view of the crystal and magnetic structures ($GxAy$) of \DSO. (c) Temperature dependence of the ordered moments. The solid line is a guide to the eye. The red dashed line indicates the AF ordering temperature at $T=3.15$~K. }
  \label{WAND}
\end{figure}

Fig.~\ref{WAND}(a) shows the NPD pattern of a crushed single crystal of \DSO\ in the paramagnetic ($T=10$~K) and magnetically ordered ($T=1.5$~K) states.
The refined unit cell parameters and atomic positions in the $Pbnm$ unit cell do agree well with previously reported data for \DSO~\cite{Liferovich}.
The AF ordering is manifested in appearance of magnetic Bragg reflections below $T_{\mathrm{N}}$.
The symmetry analysis and Rietveld refinement reveal that the magnetic group symmetry is $Pb^{\prime}n^{\prime}m^{\prime}$ and that the magnetic moments are oriented in the $ab$-plane [($G_xA_y$) representation].
A schematic view of the crystal and magnetic structures of \DSO\ is shown in Fig.~\ref{WAND}(b).
The temperature dependence of the ordered magnetic moments are presented in Fig.~\ref{WAND}(c).
A Rietveld refinement of the 1.5~K neutron diffraction dataset reveals that the ordered moments reach $m(G_x)=4.44(3)$~\mB\ and $m(A_y)=8.36(2)$~\mB\ (9.47(6)~\mB\ total). This corresponds to a canting angle of $\varphi=28^{\circ}$ to the $b$-axis.
This is in excellent agreement with our CEF calculation and magnetization measurements, and also consistent with earlier studies~\cite{Raekers,Bluschke}.

At temperatures above the ordering transition one can also observe additional diffuse scattering at low angles in the powder patterns, see Fig.~\ref{WAND}, indicating short range magnetic correlations and fluctuations above the phase transition. This is studied in better detail with single crystal neutron scattering data which are presented below.

\subsection{Magnetic dipole-dipole interaction}

\begin{figure}[tb]
  \includegraphics[width=0.7\linewidth]{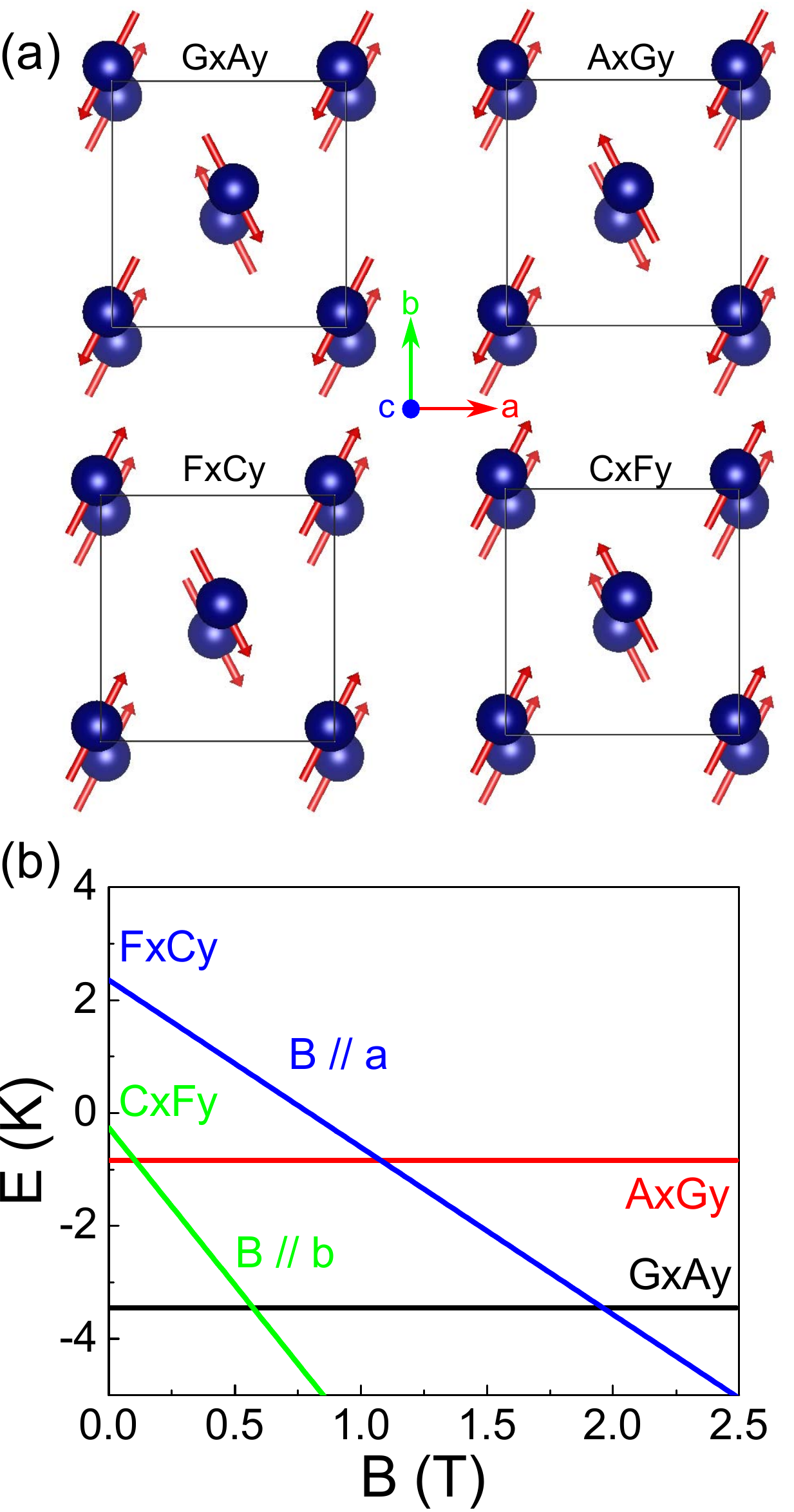}
  \caption{(a) Schematic view of the four symmetry allowed magnetic structures in \DSO. The red arrows indicate the moment directions. (b) Calculated field dependence of the dipole-dipole energy for each of the four magnetic structures, see main text. In zero field, the $GxAy$ moment configuration is calculated to have the lowest energy of about -3.4~K. For a magnetic field along the $a$-axis and $b$-axis, the $FxCy$ and $CxFy$ configurations will be a new ground state, with critical field 2~T and 0.6~T, respectively.}
  \label{dipole}
\end{figure}

In this section we discuss the magnetic ground state in the context of the magnetic dipole-dipole interaction.
We have shown that the Dy$^{3+}$ Ising moments are in the $ab$-plane along a direction with an angle $\varphi=\pm{28}^{\circ}$ relative to the $b$-axis.
According to representation analysis, four different magnetic structures ($GxAy$, $AxGy$, $FxCy$ and $CxFy$) with propagation vector $k=0$ are allowed in this case.
These four structures are shown in Fig.~\ref{dipole}(a).
Neutron diffraction shows that \DSO\ selects the $GxAy$ configuration.
Since the Dy$^{3+}$-4$f$ electrons are very localized, the 4$f$ exchange interaction is relatively weak.
On the other hand, the dipole-dipole interaction is expected to be large due to the extremely large moments in the ground state ($M_{\|}\simeq{10}$~\mB/Dy).
Since the dipole-dipole interaction is long range in nature, for each of the four magnetic structures we consider ten near neighbors in total, including eight neighbors within distance $\sim{5.7}$~{\AA} in the $ab$-plane, and two near neighbors at a distance of $\sim{4.0}$~{\AA} along the $c$-axis.
Further neighboring atoms have little influence.
The calculated dipole-dipole energies
\begin{align*}
E_{\rm dipole} =-\frac{\mu_{0}}{4\pi}\sum_{\rm i}
\frac{1}{|\mathbf{r}_{\rm i}|^3}
\left[3\left(\mathbf{m}_{0}\cdot\mathbf{\hat{r}}_{\rm i}\right)
\left(\mathbf{m}_{\rm i}\cdot\mathbf{\hat{r}}_{\rm i}\right)
-\left(\mathbf{m}_{0}\cdot\mathbf{m}_{\rm i}\right)\right]
\end{align*}
of the four magnetic structures in zero field are $GxAy=-3.61$~K, $AxGy=-0.90$~K, $CxFy=-0.26$~K, and $FxCy=2.44$~K, as presented in Fig.~\ref{dipole}(b).
As we can see, the $GxAy$ spin configuration has the lowest zero field energy compared with the non-ordered paramagnetic state and the other spin structures.
Thus, at zero field the system adopts the $GxAy$ ground state, and the AF ordering temperature $T_{\mathrm N}=3.2$ K is well consistent with the energy scale one estimates from the dipole-dipole interaction.
We also note, that the two $GxAy$ and $AxGy$ configurations would be degenerate if there were no interactions in the $ab$-plane.
This means that the in-plane interaction lifts the degeneracy and selects the ground state spin configuration.
In turn, the in-plane dipolar interaction depends on the relative tilting angle $\varphi$ of the Ising moments~\cite{Kappatsch}.
Larger value of $\varphi$ would switch the ground state from $GxAy$ to $AxGy$, as in the case of the isostructural compound YbAlO$_3$ \cite{Radha81}. Also, if we apply a magnetic field above critical along either the $a$-axis ($B_{\mathrm{crit}}\sim2$~T) or the $b$-axis ($B_{\mathrm{crit}}\sim0.6$~T), either $FxCy$ or $CxFy$ would be the new ground state,  stabilized by the Zeeman energy,
\begin{align*}
E_{\rm{Zeeman}} =-\sum_{\rm i} \mathbf{B}\cdot \mathbf{M}_{\rm{i}}/z,
\end{align*}
where $z$ is the number of ions that are summed over the magnetic unit cell. The total field dependent energy of the system is then contributed from both the dipole and Zeeman terms:
\begin{align*}
E =E_{\rm{dipole}}+E_{\rm{Zeeman}},
\end{align*}
as seen in Fig.~\ref{MBTA}(b).
Since the only difference between $GxAy$ and $CxFy$ (or $AxGy$ and $FxCy$) is the spin arrangement along the $c$-axis, the critical fields give a rough estimate of the interactions along the $c$-axis, which close the gap between the $GxAy$ and $CxFy$ (or $AxGy$ and $FxCy$) states.

It is also important to note, that the Dy-Dy dipolar interaction between two nearest neighbors along the $c$-direction is $E_{\mathrm{dipole}}^{c}\simeq{-0.82}$~K and antiferromagnetic.
This is about 2-4 times larger than the corresponding interactions in the $ab$-plane, which are ferromagnetic, $E_{\mathrm{dipole}}^{ab}\simeq{+0.45}$~K and
$E_{\mathrm{dipole}}^{ab}\simeq{+0.21}$~K (corresponding to near neighbor distances of
$3.79$~{\AA} and $4.05$~{\AA} in the plane).

\subsection{Magnetic diffuse scattering}
\label{sec_diffuse}
\begin{figure}[tb]
  \includegraphics[width=1.0\linewidth]{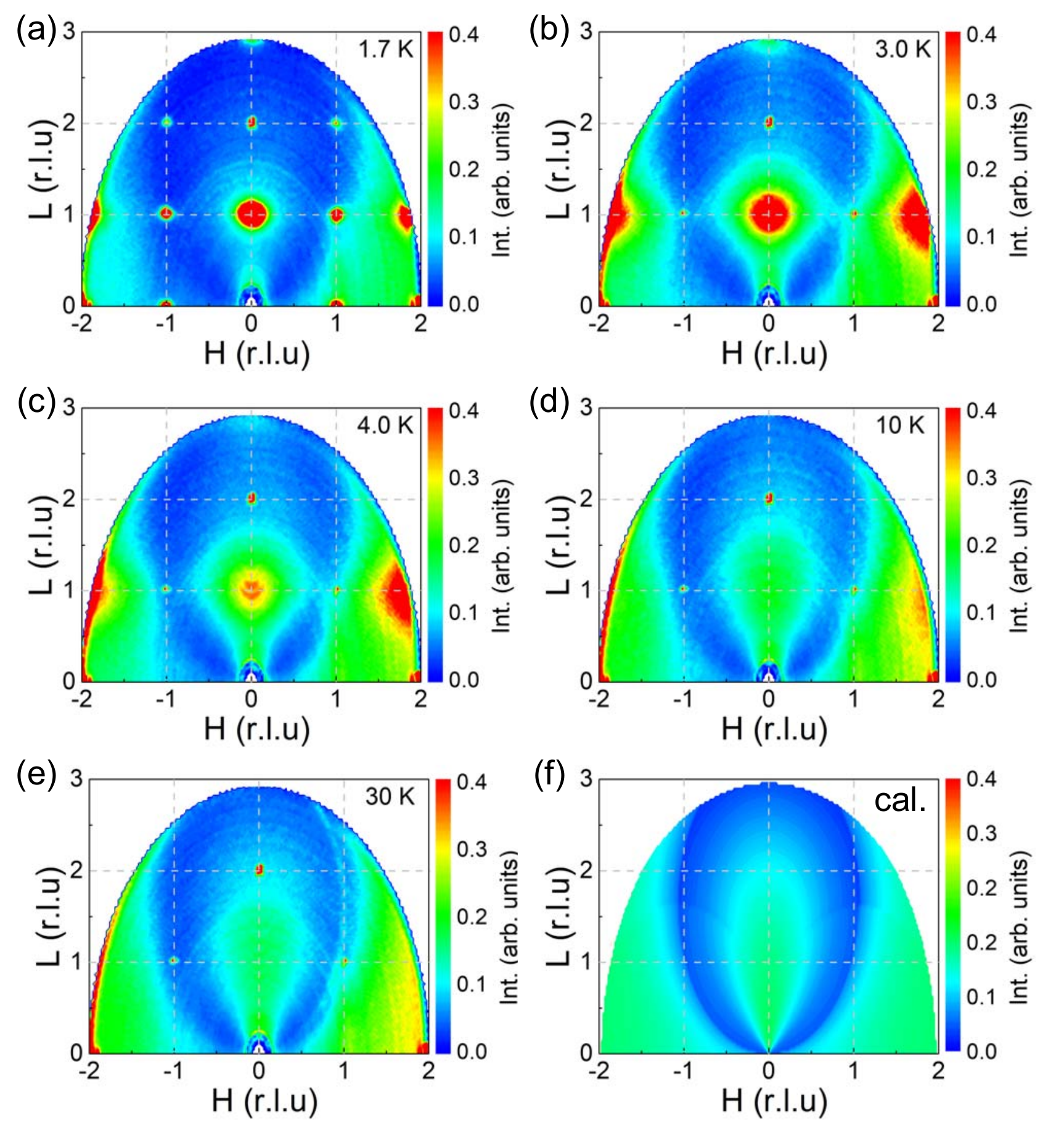}
  \caption{(a)-(e) Contour plots of the measured magnetic neutron diffuse scattering of \DSO\ in the $(H,0,L)$ plane, integrated over the energy window $E=[-0.1, 0.1]$~meV, at different temperatures 1.7~K (a), 3~K (b), 4~K (c), 10~K (d) and 30~K (e). (f) Calculated magnetic scattering factor including the absorption correction, the polarization factor and the magnetic form factor of the Dy$^{3+}$ ion. Here, r. l. u. stands for reciprocal lattice units.}
  \label{diffuse}
\end{figure}

Contour maps of the magnetic neutron scattering in the $(H,0,L)$ plane at different temperatures, above and below $T_{\mathrm{N}}$, show broad diffuse peaks near the magnetic wave vectors $Q=(0,0,1)$ and $(\pm{1},0,1)$, which are smeared out with increasing temperature (Fig.~\ref{diffuse}).
We have calculated magnetic scattering factor as:
\begin{multline}\label{Scattering_factor}
\int dE {\;} S(Q,E)\propto T(Q)\cdot|f(Q)|^{2}\cdot{S(Q)}\\
\times\left(\delta_{\alpha\beta}-\widetilde{Q}_{\alpha}\widetilde{Q}_{\beta}\right){\;}.
\end{multline}
Here, $T(Q)$ is the transmission for the \DSO\ sample,
$|f(Q)|^{2}$ is the magnetic form factor of Dy$^{3+}$,
$S(Q)$ is the magnetic structure factor, and
$\left(\delta_{\alpha\beta}-\widetilde{Q}_{\alpha}\widetilde{Q}_{\beta}\right)$ is the polarization factor.
As discussed earlier, the transverse moment component is very small, $M^2_{\perp}/M^2_{\|}\simeq{0.0025}$.
Therefore, we can safely neglect the fluctuations along the transverse direction.
Since neutrons scatter only from the component of the magnetic moment that is perpendicular to the wave vector $\mathbf{Q}$, in the present case the calculated magnetic polarization factor is
\begin{equation}
\label{polarization_factor_equation1}
\delta_{\alpha\beta}-\widetilde{Q}_{\alpha}\widetilde{Q}_{\beta}=
1-\frac{Q^2_{\rm{h}}\cdot\sin^{2}\varphi}{Q^2_{\rm{h}}+Q^2_{\rm{l}}}=
\frac{Q^2_{\rm h}\cdot\cos^{2}\varphi+Q^2_{\rm l}}{Q^2_{\rm h}+Q^2_{\rm l}}{\;},
\end{equation}
where $\varphi=\pm{28}^{\circ}$, $Q_{\rm{h}}=2\pi h/a$ and $Q_{\rm{l}}=2\pi l/c$.
The neutron transmission factor was explicitly included because of the strong absorption by Dy and Sc. The calculated 1/e absorption length for a 3.32~meV neutron is only about 0.215~mm~\cite{Patra}.
The calculated  magnetic  scattering factor (\ref{Scattering_factor}) is shown in Fig.~\ref{diffuse}(f).
Good agreement was found with the data taken at 10~K (Fig.~\ref{diffuse}(d)) and 30~K (Fig.~\ref{diffuse}(e)).
The dip like feature that went through wave vector $Q=(\pm1,0,1)$ is due to strong neutron absorption along the plate.
We have adopted the lattice Lorentzian function~\cite{Igor2015} to describe the magnetic structure factor:
\begin{multline}
\label{lattice_lorentzian}
S(Q)\propto\frac{\sinh(c/\xi_{l})}{\cosh(c/\xi_{l})-\cos(\pi(l-1))}\\
\times\frac{\sinh(a/\xi_{h})}{\cosh(a/\xi_{h})-\cos(\pi{h})}.
\end{multline}
Here, $\xi_{l}$ and $\xi_{h}$ are the correlation lengths along the $c$ and $a$-axes in units~\AA.

\begin{figure}[tb]
  \includegraphics[width=0.7\linewidth]{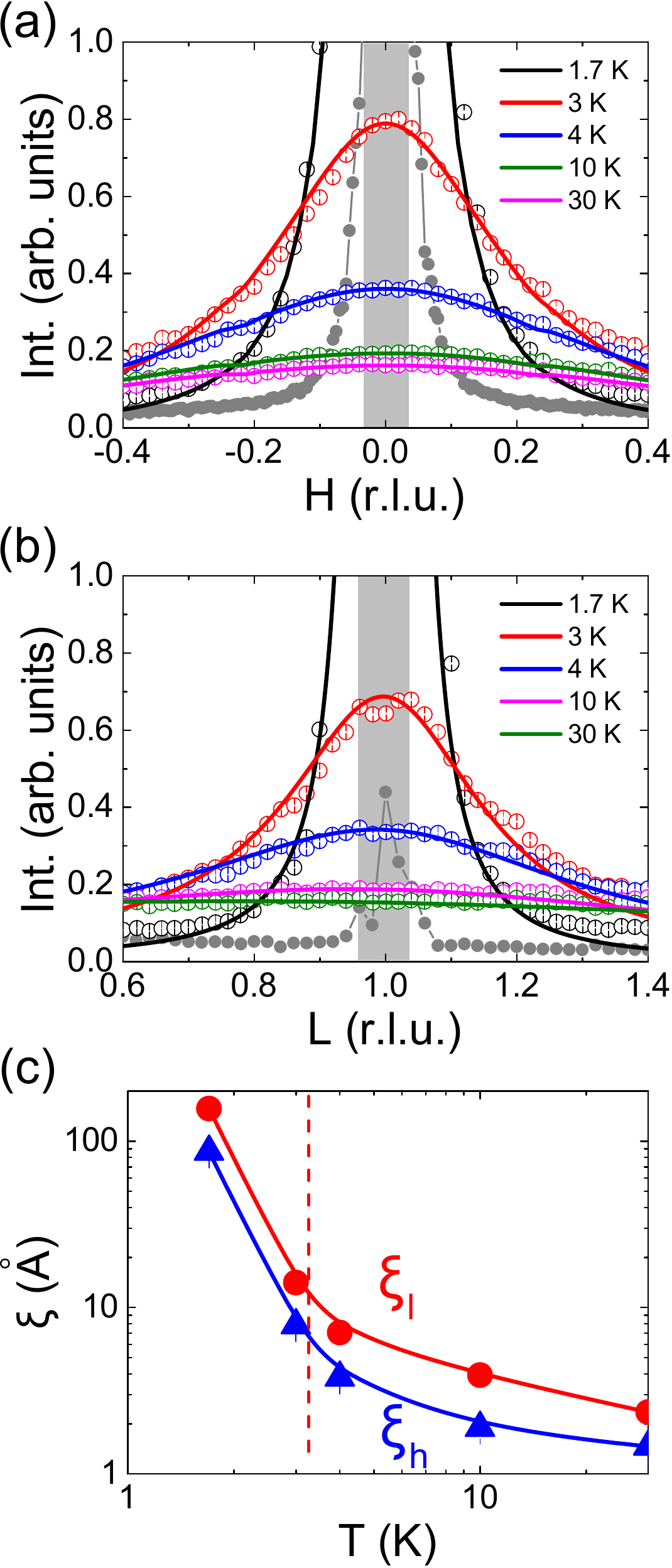}
  \caption{(a) Cut along wave vector $Q_{\rm{h}}$, integrated over $E=[-0.1,0.1]$~meV,
  $Q_{\rm{l}}=[0.9,1.1]$, and $Q_{\rm{k}}=[-0.2,0.2]$, measured at different temperatures as indicated. (b) Cut along wave vector $Q_{\rm{l}}$, integrated over $E=[-0.1,0.1]$ meV,
  $Q_{\rm{h}}=[-0.1,0.1]$, and $Q_{\rm{k}}=[-0.2,0.2]$, measured at different temperatures as indicated. The instrumental resolution (gray bar) is estimated from a cut through the pure nuclear (002) peak (gray solid circles). (c) The temperature dependent correlation length from the fit of the lattice Lorentzian functions, as explained in the text. The red dashed line indicates the AF ordering temperature.}
  \label{Correlation}
\end{figure}

We show constant energy cuts along wave vector $Q_{\rm{h}}$ and $Q_{\rm{l}}$, integrated over $E=[-0.1,0.1]$~meV, $Q_{\rm{l}}=[0.9,1.1]$, and $Q_{\rm{k}}=[-0.2,0.2]$ in Fig.~\ref{Correlation}(a) and (b).
The fits of the lattice Lorentzian function are shown as solid lines.
The temperature dependent correlation lengths from the fits are shown in Fig.~\ref{Correlation}(c).
At temperatures 30~K and 10~K both correlations lengths, along $Q_{\rm{h}}$ and $Q_{\rm{l}}$, are short, much less than one lattice unit.
However, as approaching the magnetic ordering temperature, significant correlations build up already at 4~K along both directions, which are greatly enhanced below the ordering temperature.
A slight anisotropy was observed between wave vector $Q_{\rm{h}}$ and $Q_{\rm{l}}$, where the correlation along $Q_{\rm{l}}$ is always larger than that along $Q_{\rm{h}}$, as expected from the estimate of the  $E_{\mathrm{dipole}}^{c} / E_{\mathrm{dipole}}^{ab}$ dipolar interactions.

\section{Discussion and Conclusion}

In summary, the low temperature magnetic properties of \DSO\ hasve been well characterized through a combination of CEF calculations, magnetization, powder and single crystal neutron diffraction measurements.
All our results are consistent with an Ising Dy$^{3+}$ ground state doublet with wave function $\lvert\pm{15}/2\rangle$, where the local easy axis makes an angle of $\varphi=28^{\circ}$ with the $b$-axis.
The Dy$^{3+}$ Ising moments order magnetically at low temperatures ($T_{\rm{N}}=3.15(5)$~K), and the AF magnetic ground state $GxAy$ is selected mainly by the dipole-dipole interaction.

Strong magnetic diffuse scattering is observed for temperatures up to 30~K, which indicates the persistence of strong critical fluctuations over a wide temperature range. This is rather surprising. Considering that the Ising like moments are rather large,
$M_{\|}\simeq{10}$~\mB, one would usually expect to see a clean mean-field like transition with very weak critical fluctuation~\cite{Wu2011}.
Critical fluctuations could be enhanced by the low dimensionality, as seen in other rare-earth based one-dimensional systems~\cite{Wu2016,Miiller}.

No inelastic excitations have been observed at all in \DSO\ below the first excited CEF level.
To understand this puzzling observation, one may go back to the ground state wave function.
First, one notes that any inelastic scattering due to transverse fluctuations should be strongly suppressed.
Because of the strong Ising like single ion anisotropy, which can be quantified as
$M_{\parallel}/M_{\perp}=\langle{E}_{\rm{0}\pm}|J_{\rm{z'}}|{E}_{\rm{0}\pm}\rangle/
\langle{E}_{\rm{0}\pm}|J_{\rm{y'}}|{E}_{\rm{0}\pm}\rangle\simeq{20}$,
the neutron scattering intensity due to transverse fluctuations is expected to be about two orders of magnitude smaller than the scattering intensity from longitudinal fluctuations,  $M^2_{\parallel}/M^{2}_{\perp}\simeq{400}$,
at temperatures and fields smaller the the first CEF excited level $\Delta_{1}$.
Thus, any fluctuations along the transverse direction, such as spin waves, would be extremely weak.
One would then expect to see strong fluctuations along the longitudinal directions from the Dy$^{3+}$ Ising moments.
In the most general case the ground state wave function can be expressed as
\begin{multline}
\label{E0_general}
E_{\rm{0}\pm}=\alpha\lvert\pm{15}/2\rangle
+\beta\lvert\pm{13}/2\rangle
+\gamma\lvert\pm{11}/2\rangle\\
+\delta\lvert\pm{7}/2\rangle
+\epsilon\lvert\pm{5}/2\rangle
+\zeta\lvert\pm{3}/2\rangle
+\eta\lvert\pm{1}/2\rangle\\
+\alpha'\lvert\mp{15}/2\rangle
+\beta'\lvert\mp{13}/2\rangle
+\gamma'\lvert\mp{11}/2\rangle\\
+\delta'\lvert\mp{7}/2\rangle
+\epsilon'\lvert\mp{5}/2\rangle
+\zeta'\lvert\mp{3}/2\rangle
+\eta'\lvert\mp{1}/2\rangle{\;}.
\end{multline}
Comparing to the calculated ground state wave function, see above, we see that most of these contributions are missing ($\eta=\alpha'=\beta'=\gamma'=\delta'=\epsilon'=\zeta'=\eta'=0$) due to the constraint of the point site symmetry.
Therefore, any Hamiltonian matrix element, which connects the 'up' and 'down' states of moment $\langle{E}_{\rm{0}\mp}|S^{+}, S^{-}|{E}_{\rm{0}\pm}\rangle=\alpha\beta'+\alpha'\beta+...=0$ is absent.
This indicates that neutrons can neither flip the Dy$^{3+}$ ground Ising moments nor propagate them, due to the selection rule $\Delta{S}=1$.
In other words, in the case of \DSO, these possible spinon excitations along the longitudinal directions are 'hidden' to neutrons.
It is interesting to notice that a splitting of ground-state was observed in the optical absorption spectra of the related DyAlO$_3$ compound~\cite{Schuchert}, which was claimed to originate from the one-dimensional interactions along the $c$-axis.
Further studies using other direct or indirect techniques, such as ac-susceptibility, or optical absorption, for which such selection rules don't apply, would be needed to reveal the 'hidden' low-energy dynamics.

\begin{acknowledgments}
We thank J.~M. Sheng and Q. Zhang for the help with the neutron data refinement.
We would like to thank A. Christianson, I. Zaliznyak, M. Mourigal, Z.~T. Wang, and C. Batista for useful discussions.
The research at the Spallation Neutron Source (ORNL) is supported by the Scientific User Facilities Division, Office of Basic Energy Sciences, U.S. Department of Energy (DOE).
Research supported in part by the Laboratory Directed Research and Development Program of Oak Ridge National Laboratory, managed by UT-Battelle, LLC, for the U.S. Department of Energy.
This work was partly supported by the U.S. Department of Energy (DOE), Office of Science, Basic Energy Sciences (BES), Materials Science and Engineering Division.
\end{acknowledgments}

\appendix
\label{append}

\section{Point charge model calculations}
A point charge model considers crystalline electric field effects as a perturbation to the appropriate free-ion $4f$ wave functions and energy levels and the perturbed crystalline potential energy may be rewritten as:

\begin{align}
\label{MvsB}
W_c = \sum_{i}q_{i}V_{i} = \sum_{i}\sum_{j} \frac{q_{i}q_{j}}{\lvert R_{j}-r_{i} \rvert}
\end{align}

where $q_{i}$ and $q_{j}$ are charges of magnetic ions and ligand ions, $\sum_{j}$ is the  sum over all neighbor charges~\cite{Hutchings} and $\lvert R_{j}-R_{i} \rvert$ is the distance between magnetic ion and ligand charge. The calculated first excited state is

\begin{multline*}
\label{E1}
\lvert{E}_{\pm}\rangle_{1}=-0.09\lvert\pm{15}/2\rangle
\mp{0.976}\lvert\pm{13}/2\rangle\\
+0.191\lvert\pm 11/2\rangle
\pm{0.017}\lvert\pm{9}/2\rangle
+0.029\lvert\pm{7}/2\rangle\\
\mp{0.011}\lvert\pm{5}/2\rangle
+0.038\lvert\pm {3}/2\rangle
\mp{0.007}\lvert\pm {1}/2\rangle\\
+0.018\lvert\mp{1}/2\rangle
\mp{0.003}\lvert\mp{3}/2\rangle
+0.009\lvert\mp{5}/2\rangle\\
+0.003\lvert\mp{9}/2\rangle
\pm{0.002}\lvert\mp {11}/2\rangle{\;}.
\end{multline*}

The calculated CEF parameters from the point charge model and resulting the energy levels are shown in Table~\ref{Blm_table} and ~\ref{tab:CEF} below, respectively.
\begin{table}[h]
\caption {Set of $B_{l}^{m}$ (meV) parameters calculated from the point charge model.}
\label{Blm_table}
\begin{tabular}{ l @{\qquad} @{\qquad}c }
\toprule
 $B_2^0$ & $-4.35\times10^{-1}$\\
 $B_2^1$ & $0.45\times10^{-1}$\\
 $B_2^2$ & $-4.03\times10^{-1}$\\
 \hline
 $B_4^0$ & $-0.22\times10^{-3}$\\
 $B_4^1$ & $-0.14\times10^{-3}$\\
 $B_4^2$ & $3.2\times10^{-3}$\\
 $B_4^3$ & $2.5\times10^{-3}$\\
 $B_4^4$ & $-1.0\times10^{-3}$\\
 \hline
 $B_6^0$ & $0.02\times10^{-5}$\\
 $B_6^1$ & $-0.32\times10^{-5}$\\
 $B_6^2$ & $0.08\times10^{-5}$\\
 $B_6^3$ & $-1.13\times10^{-5}$\\
 $B_6^4$ & $0.16\times10^{-5}$\\
 $B_6^5$ & $3.86\times10^{-5}$\\
 $B_6^6$ & $0.32\times10^{-5}$\\
\botrule
\end{tabular}
\end{table}

\begin{table}[h]
\caption{The energy levels and out of ground state transition probabilities.
\label{tab:CEF}}
\begin{ruledtabular}
\begin{tabular}{lll}
 & $E$~(meV)  & $<n| \mathrm{J}_{\perp} |m>^2$  \\
\hline
$|E_0\rangle \rightarrow |E_1\rangle$& 22.9 & 8.2 \\
$|E_0\rangle \rightarrow |E_2\rangle$& 37.6 & 0.1 \\
$|E_0\rangle \rightarrow |E_3\rangle$& 43.1 & 0.02 \\
$|E_0\rangle \rightarrow |E_4\rangle$& 52.2 & 0.05 \\
$|E_0\rangle \rightarrow |E_5\rangle$& 63.3 & 0.02 \\
$|E_0\rangle \rightarrow |E_6\rangle$& 79.7 & 0 \\
$|E_0\rangle \rightarrow |E_7\rangle$& 98.1 & 0 \\
\end{tabular}
\end{ruledtabular}
\end{table}

\pagebreak

%

\end{document}